\documentclass[12pt]{article}
\usepackage{pic04}
\usepackage{hyperref}
\usepackage{url}
\usepackage{graphicx}

\begin{document}

\title{\bf STRUCTURE FUNCTION MEASUREMENTS AND POLARISED CROSS SECTION
MEASUREMENTS FROM HERA}
\author{
Andrew Mehta   \\
{\em Oliver Lodge Lab,
  Dept. of Physics,
  Liverpool University,
  Liverpool,
  L69 7ZR,
  UK }}
\maketitle

%
%
%
%
%
%
\vspace{4.5cm}
%

\baselineskip=14.5pt
\begin{abstract}
  Recent measurements of inclusive and semi-inclusive measurements
  from the HERA collaborations are presented. The measurements include
  neutral current structure functions $F_2$, $F_L$ and $xF_3$; the
  charged current cross section including first measurements of the
  dependence on electron polarisation; and measurements of the heavy
  quark structure functions $F_2^{c\bar{c}}$ and $F_2^{b\bar{b}}$.
\end{abstract}
\newpage

\baselineskip=17pt

\section{Introduction}
\label{intrp}
Deep inelastic scattering (DIS) has long been used to measure the
structure of hadrons. Measurements with a fixed target helped
establish the quark parton model and the theory of the strong force,
quantum chromodynamics (QCD). HERA, being the first colliding beam DIS
experiment, enables proton structure to be measured in hitherto
unexplored regions. The measurements of proton structure at HERA,
particularly at very low fractional parton momentum, have yielded
extremely precise parton distribution functions (PDFs), essential for
making QCD predictions at present and future hadron colliders.

In the years 1998--2000 HERA was operated in both $e^+p$ and $e^-p$
scattering modes at a centre of mass energy of $\sqrt{s}=320$~GeV. The
large data samples collected has allowed a determination of all the
possible neutral current (NC, $ep \rightarrow eX$) structure functions
$F_2$, $F_L$ and $xF_3$ for the first time at HERA. The structure
function $F_2$ is sensitive to all quark species and dominates the
cross section throughout the accessible phase space.  The quantity
$xF_3$ is sensitive to the valence quarks.  Since the cross section
only becomes sensitive to $xF_3$ via the exchange of the $Z^0$ boson,
its influence is limited to the very high $Q^2$ electroweak regime.
Finally, $F_L$ is sensitive to higher order gluon radiation processes
providing valuable confirmation of the gluon content of the proton.

Like the neutral current cross section, the charged current (CC , $ep
\rightarrow \nu X$) cross section is an important tool to measure the
structure of the proton, particularly since the charged current
process is sensitive to the quark flavour decomposition of the proton.
Measurements of the $e^+p$ and $e^-p$ scattering cross sections have
allowed independent determination of the $u$ quark and $d$ densities.
In the years 2003-2004 HERA operated in $e^+p$ mode with longitudinally
polarised $e^+$ beam, allowing the dependence of the CC cross section
on polarisation to be measured for the first time.

The structure functions are usually presented in terms of the
kinematic variables Bjorken $x$, the fraction of the proton's momentum
carried by the struck quark and $Q^2$ the negative square of the
4-momentum transfer of the exchanged boson. $Q^2$ may be interpreted
as the resolving power of the exchange, with increasing $Q^2$ able to
resolve smaller distances within the proton. They may be derived from
the differential cross sections as

\begin{equation} 
 \frac{{\rm d} \sigma_{NC}(e^\pm p)}{{\rm d} x {\rm d}Q^2}\simeq {\frac{2 \pi \alpha^2}{xQ^4}} [Y_+ \tilde{F}_2 {\mp} Y_- x\tilde{F}_3 -y^2 \tilde{F}_L]
\label{eqn:sf}
\end{equation} 

Here $\alpha$ is the electromagnetic coupling constants and $Y_\pm =1
\pm(1-y)^2$ where $y=Q^2/sx$. A similar relationship holds for CC
interactions~\cite{nccctheory}.

\section{Proton Structure at Low $x$}
\label{lowx}
\begin{figure}[h]
\includegraphics[width=11.5cm]{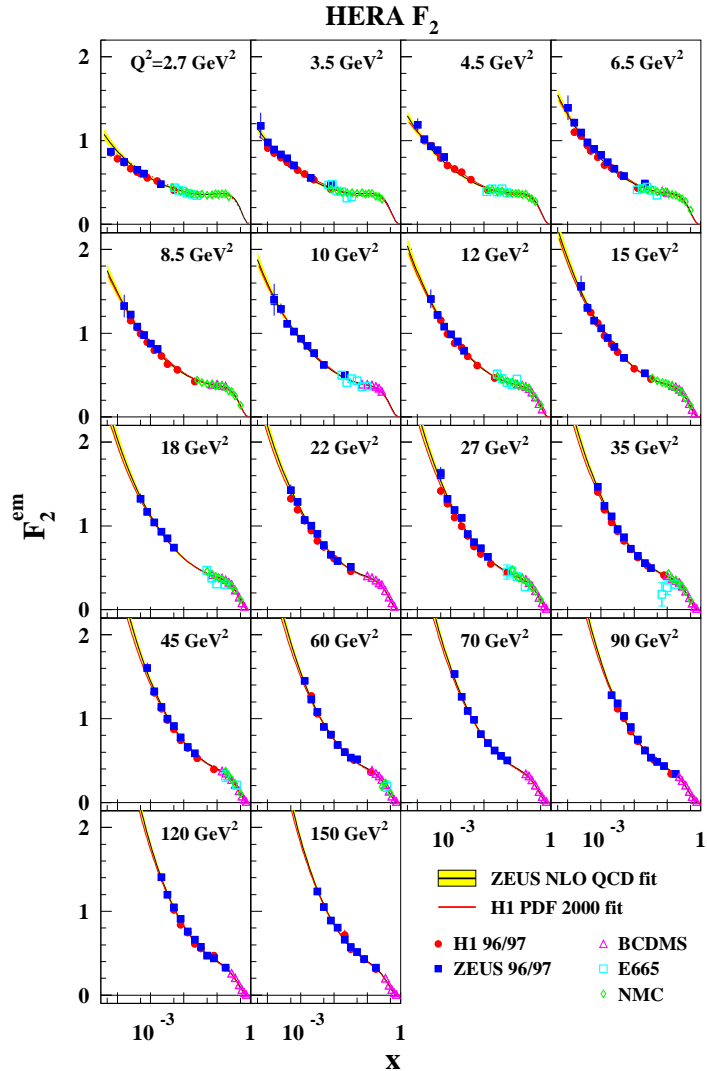}
 \caption{\it
   Measurements of the structure function $F_2$ from the HERA
   collaborations and fixed target experiments. The data are plotted
   as a function of $x$ for various fixed values of $Q^2$.  Also
   included are the results of a NLO QCD fit to the data.
\label{fig:f2x}
 }
\end{figure}

The large integrated luminosity achieved by HERA in the last few years
has provided  data samples of several million events for NC
measurements at low $x$ and low $Q^2$. This has allowed the
determination of the proton structure function $F_2$ to an accuracy of
$\simeq 2\%$ \cite{lowq2papers1,lowq2papers2}.  Example measurements
of $F_2$ from HERA and fixed target experiments are shown in
figure~\ref{fig:f2x}.  $F_2$, which is sensitive to the total quark
density of the proton, is seen to rise steeply as $x$ decreases.  The
data are compared to a next to leading order (NLO)
quantum-chromodynamics (QCD) fit, which describes all data for
$Q^2>3$~GeV very well.

\begin{figure}[htb]
\includegraphics[width=8cm]{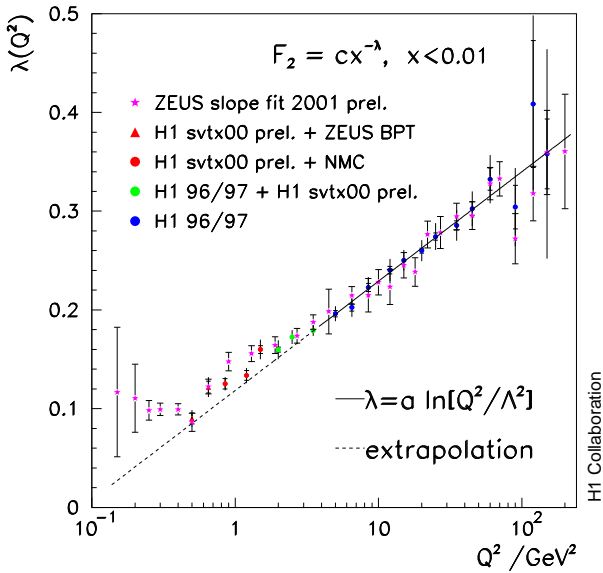}
\caption{ Measurements of the slope parameter $\lambda$ where the $F_2$ is parameterised as  $F_2=c(Q^2) x ^{-\lambda}$. }
\label{fig:lammergez}
\end{figure}

It may also be seen in figure~\ref{fig:f2x} that the rise towards
lower $x$ becomes more pronounced as $Q^2$ increases. In order to
explore this feature more quantitatively $F_2$ is parameterised as
$F_2=c(Q^2) x ^{-\lambda}$ and the slope parameter $\lambda$ is
plotted as a function of $Q^2$ in figure~\ref{fig:lammergez}. For
$Q^2>1$~GeV it can be seen that $\lambda$ increases. This can be
explained by the fact that as $Q^2$ increases the resolving power of
the probe increases and more and more splitting processes of the form
$g \rightarrow q \bar{q}$ are resolved. The components after the
spitting carry less momentum and so are observed at lower $x$. At
$Q^2<1$ GeV the data tend to reach a constant value for $\lambda$. In
this region the data are difficult to describe using perturbative QCD
although non-perturbative models do have some
success~\cite{verylowq2papers}.

\begin{figure}[htbp]
\centerline{\hbox{ \hspace{0.2cm}
    \includegraphics[width=6.5cm]{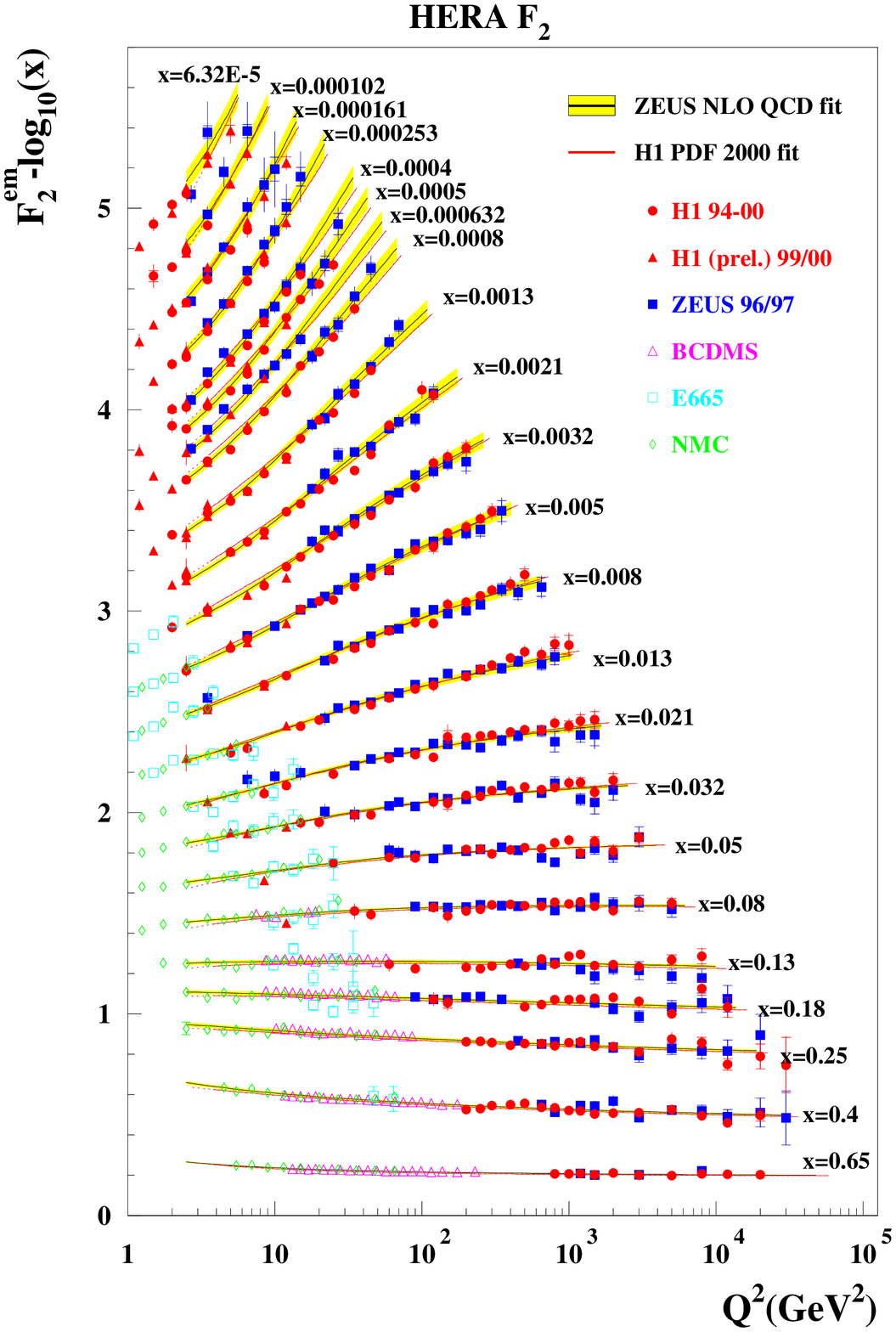}
    \hspace{0.3cm}
    \includegraphics[width=6.5cm]{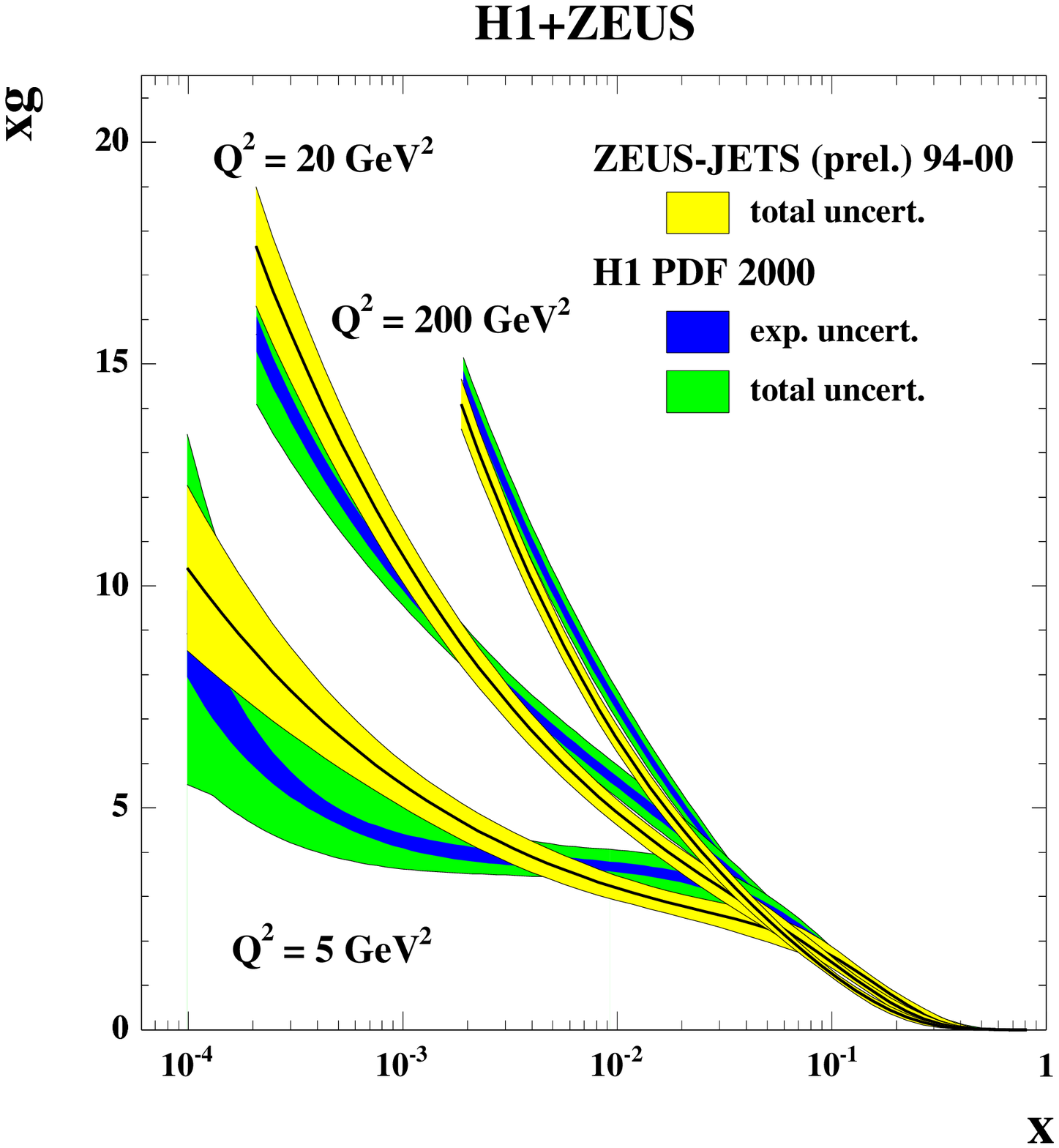}
    }
  }
\caption{  (a) Measurements of the structure function $F_2$ from 
  the HERA collaborations and fixed target experiments. The data are
  plotted as a function of $Q^2$ for various fixed values of $x$.
  Also included is the results of a NLO QCD fit to the data.  (b) The
  gluon density derived from the NLO QCD fit, plotted as a function of
  $x$ for various $Q^2$}
\label{fig:f2andglu}

\end{figure}

It is also interesting to plot $F_2$ as a function of $Q^2$ at fixed
$x$ as shown in figure~\ref{fig:f2andglu}.(a). If the proton were made
solely of 3 quarks $F_2$ should remain constant or scale with $Q^2$.
What is seen is negative scaling violations at large $x$, approximate
scaling at $x \simeq 0.2$ and very large positive scaling violations
at low $x$. In QCD these scaling violations are interpreted as gluon
radiation. As $Q^2$ increases there is an increasing chance that a
valence quark radiates a gluon and so decreasing the quark density at
higher $x$. Conversely at lower $x$ there is an increased chance of a
radiated gluon splitting into a quark pair so the density increases.

NLO QCD theory shows that the gluon density is related to the
differential of $F_2$: ${\rm d}F_2 / {\rm d}Q^2 \propto xg$.
Such a determination has been performed and the results are
shown in figure~\ref{fig:f2andglu}.(b). Similar scaling violations
are observed to the quark density of the proton.
\section{Determination of the longitudinal structure function $F_L$ }

\begin{figure}[htbp]
\includegraphics[width=10cm]{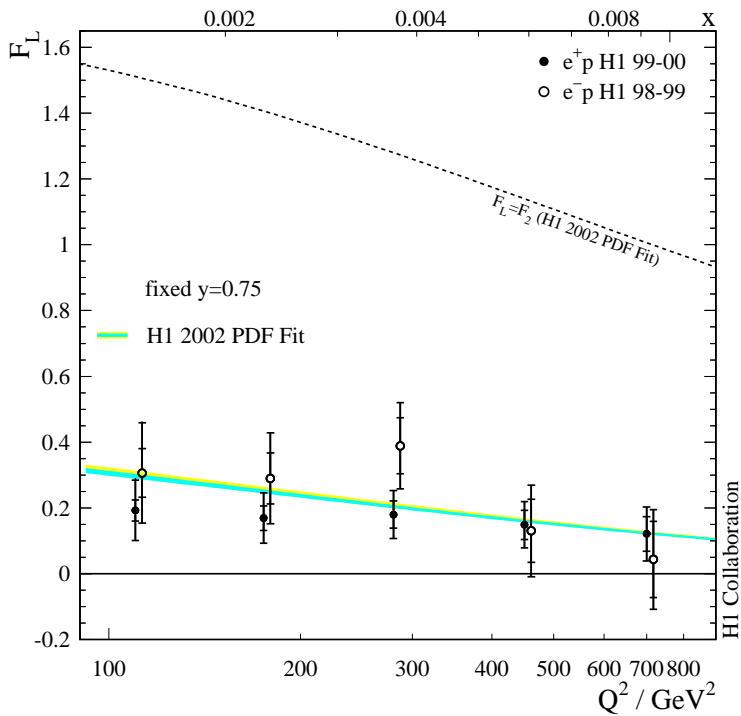}
\caption{Determination of the structure function $F_L$. The data are
  plotted as a function of $Q^2$ and $x$ for a fixed values of $y$.
  Measurements from the $e^+p$ data and the $e-p$ data are shown.  The
  data are compared to the results of the NLO QCD fit. The 2 extreme
  possibilities $F_L=0$ and $F_L=F_2$ are also shown.}
\label{fig:fl}

\end{figure}

The structure function $F_L$ has not been directly determined at HERA
since a run taken at lower beam energies has not yet been performed.
The data, however, are sensitive to $F_L$ at high $y$ as can be seen
by examining equation~\ref{eqn:sf}.  H1 have used the approach that
since $F_2$ is well determined over a wide range of $x$ and $Q^2$ it
may be extrapolated into the region at high $y$ using the QCD fit
\cite{lowq2papers1,nccch1}. Thus, by subtracting a term that depends
on $F_2$ and making a very small correction for $xF_3$ one may
determine $F_L$.  Such a determination is shown in figure~\ref{fig:fl}
for various different $Q^2$ and $x$ for a fixed $y=0.75$.
Measurements from the $e^+p$ data and the $e^-p$ data are shown. It
can be seen that the data are inconsistent with either $F_L=0$ or
$F_L=F_2$. The NLO QCD fit shows good agreement to the data. This is
an important test of QCD since $F_L$ only arises through higher order
corrections. Measurements have also been made at lower $Q^2$
\cite{lowq2papers1} which also show agreement to the theory. 

ZEUS have made a direct measurement of $F_L$ using events where the
incoming $e^+$ radiates a photon, so decreasing the centre of mass
collision energy. Although the errors on the first measurement are
large it shows that the method works and reduced errors may be
possible with high luminosity HERA II running.

\section{Neutral and Charged Current Cross Sections at high $Q^2$}
\label{nc}

\begin{figure}[htbp]
\includegraphics[width=13cm]{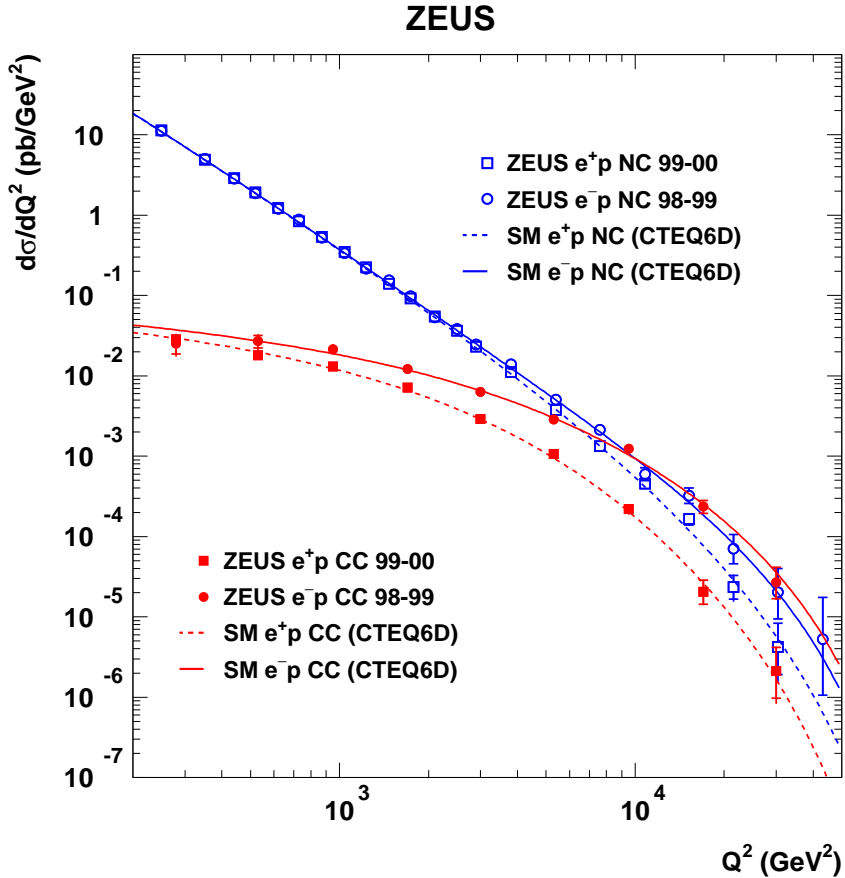}
\caption{ The $Q^2$ dependences of the NC (circles) and CC (squares) cross
  sections $ {\rm d} \sigma/{\rm d} Q^2$ are shown for the combined
  $94-00$ $e^+p$ and $98-99$ $e^-p$ measurements.  The data are
  compared to the Standard Model expectations determined from a NLO
  QCD fit. }
\label{fig:hiq2a}

\end{figure}

The NC and CC single differential cross sections ${\rm d} \sigma/{\rm
  d}Q^2$ at high $Q^2$ are shown in fig.~\ref{fig:hiq2a} for both
$e^+p$ and $e^-p$ scattering \cite{nccch1,nccch1e-,nccczeus}. 
The NC data are seen to fall with the typical $1/Q^4$ behaviour
as expected for a photon propagator. The CC cross section is suppressed
at low $Q^2$ due to the large $W$ boson mass.  At $Q^2 > M_W^2$
the cross sections become comparable as expected from the unification
of the electroweak force. The measurements are compared to a NLO
QCD fit which provides a good description of the data.

\begin{figure}[htbp]
\setlength{\unitlength}{1 mm}
\includegraphics[width=13cm]{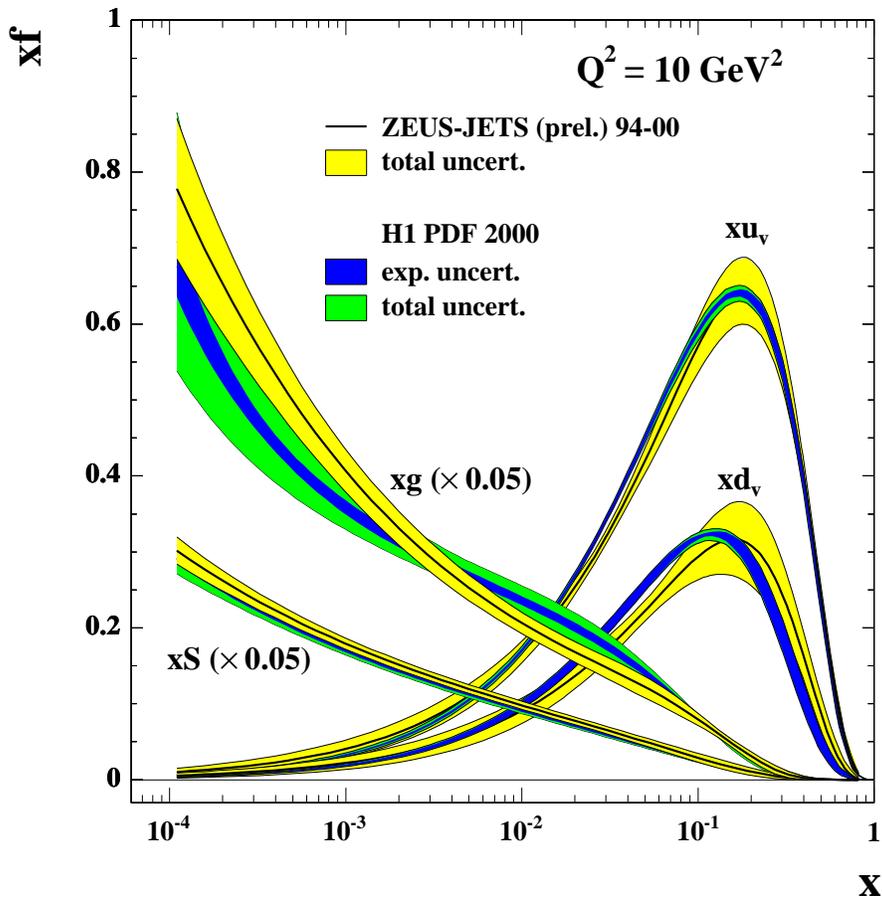}
\caption{
  The parton distributions as determined from the H1 PDF 2000 fit to
  H1 inclusive data only and those from the ZEUS-JETS fit to ZEUS
  inclusive and jet data. The distributions are shown at
  $Q^2=10$~GeV$^2$.}
\label{fig:hiq2b}

\end{figure}

The combination of four cross sections, NC and CC in both $e^+p$ and
$e^-p$ scattering, allows the flavour separation of the proton to be
achieved with minimal assumptions using HERA data alone for the first
time. NLO QCD fits have been made by both collaborations to this data
taking into account experimental and theoretical uncertainties
\cite{nccch1,zeusfit}.  ZEUS include their jet data in order to
reduce uncertainties particularly in the medium x region ($x simeq
0.1$), where the HERA inclusive data do not constrain the gluon very
precisely. The results of these fits are shown in fig.~\ref{fig:hiq2b}
for the $u$ and $d$ valence quark densities $xu_v$, $xd_v$; the sea
quark density $xS$ and the gluon density $xg$. The results of the two
fits agree well.

The cross section measurements has been improved through a reduction
of the systematic uncertainties which dominate the NC measurement up
to $Q^2\simeq 1\,000$~GeV$^2$. The double differential NC cross
section now has a total systematic uncertainty of typically about
$3\%$ compared to $6\%$ previously.  This reduction in the systematic
uncertainty has allowed the $u$ density to be determined to a
precision of typically $3\%$, and the $d$ density with a precision of
$10\%$ at $x=0.4$.

\section{Measurement of the structure function $xF_3$}
\begin{figure}[htbp]
\includegraphics[width=13cm]{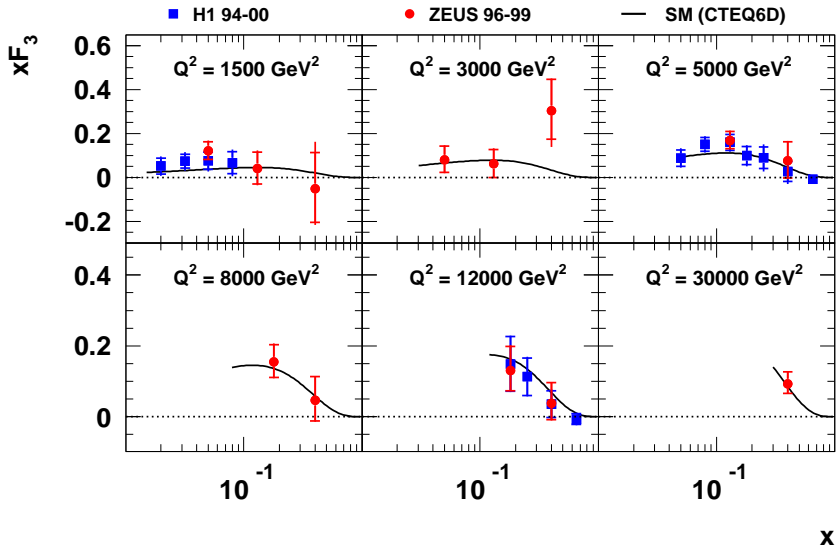}
\caption{ Measurements of the structure function $xF_3$. The data are
  plotted as a function of $x$ for various fixed values of $Q^2$.
  Also included is the results of a NLO QCD fit to the data.}
\label{fig:xf3}
\end{figure}

From equation~\ref{eqn:sf} it can be seen that the structure function
$xF_3$ may be measured by subtracting the $e^+p$ NC data from the
$e^-p$ data. Although $xF_3$ only becomes a non-negligible part of the
cross section at large $Q^2$ when $Q^2 \simeq M_Z^2$ there are now
enough statistics for a first measurement from HERA. This is shown in
figure~\ref{fig:xf3}. The structure function $xF_3$ is particularly
important because it is only sensitive to the valence quarks. The
measurements from HERA show good agreement to the NLO QCD fit. The
excess seen at low $x$ is not significant due to the relatively large
errors on the data at the moment.

\section{Polarised CC cross section}
\begin{figure}[htbp]
  \includegraphics[width=13cm]{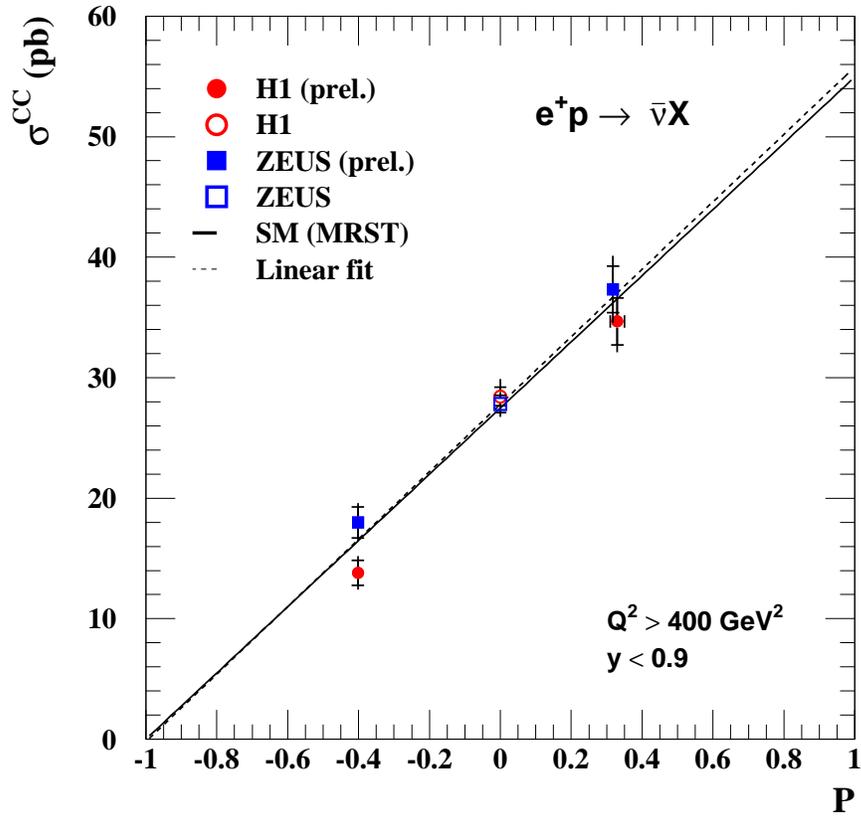}
\caption{Measurements of the CC cross section as a function of
  the longitudinal polarisation of the the $e^+$ beam.}
\label{fig:polcc}
\end{figure}

The HERA upgrade included installation of spin rotators in the HERA
ring.  This enables longitudinal polarisation of the $e^+$ beam. The
first measurements \cite{polcc} of the CC cross section as a function of
polarisation ($P$) are shown in figure~\ref{fig:polcc}. It is seen that
the cross section is smaller for negative polarisation than for
positive.  In the Standard Model, due to the absence of right handed
charged currents, it is expected that the CC cross section follows a
linear dependence, with zero expectation at $P=-1$. The Standard Model
agrees well with the data.

\section{Measurement of $F_2^{c\bar{c}}$ and $F_2^{b\bar{b}}$}
\begin{figure}[htbp]
  \includegraphics[width=12cm]{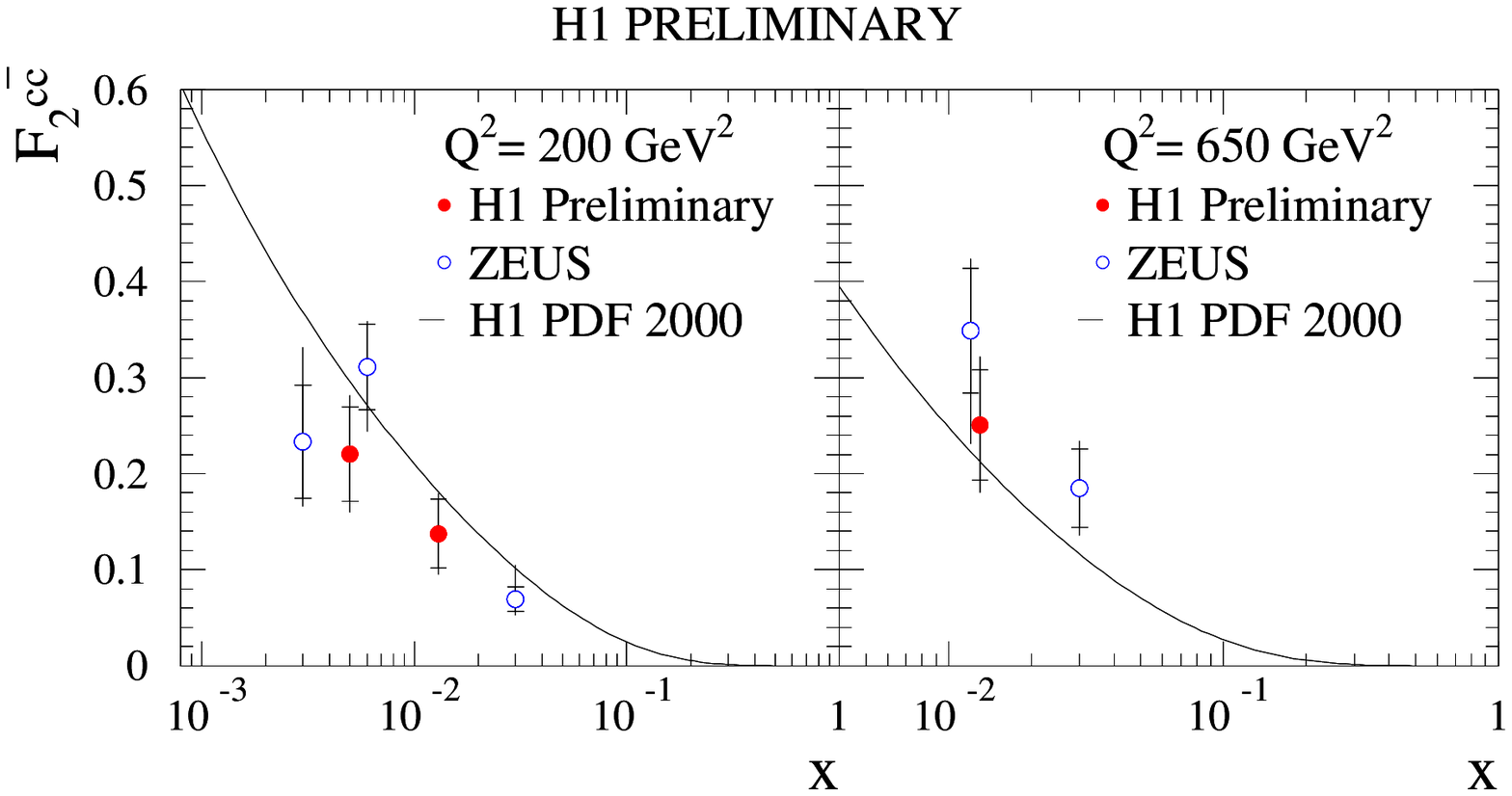}
  \includegraphics[width=12cm]{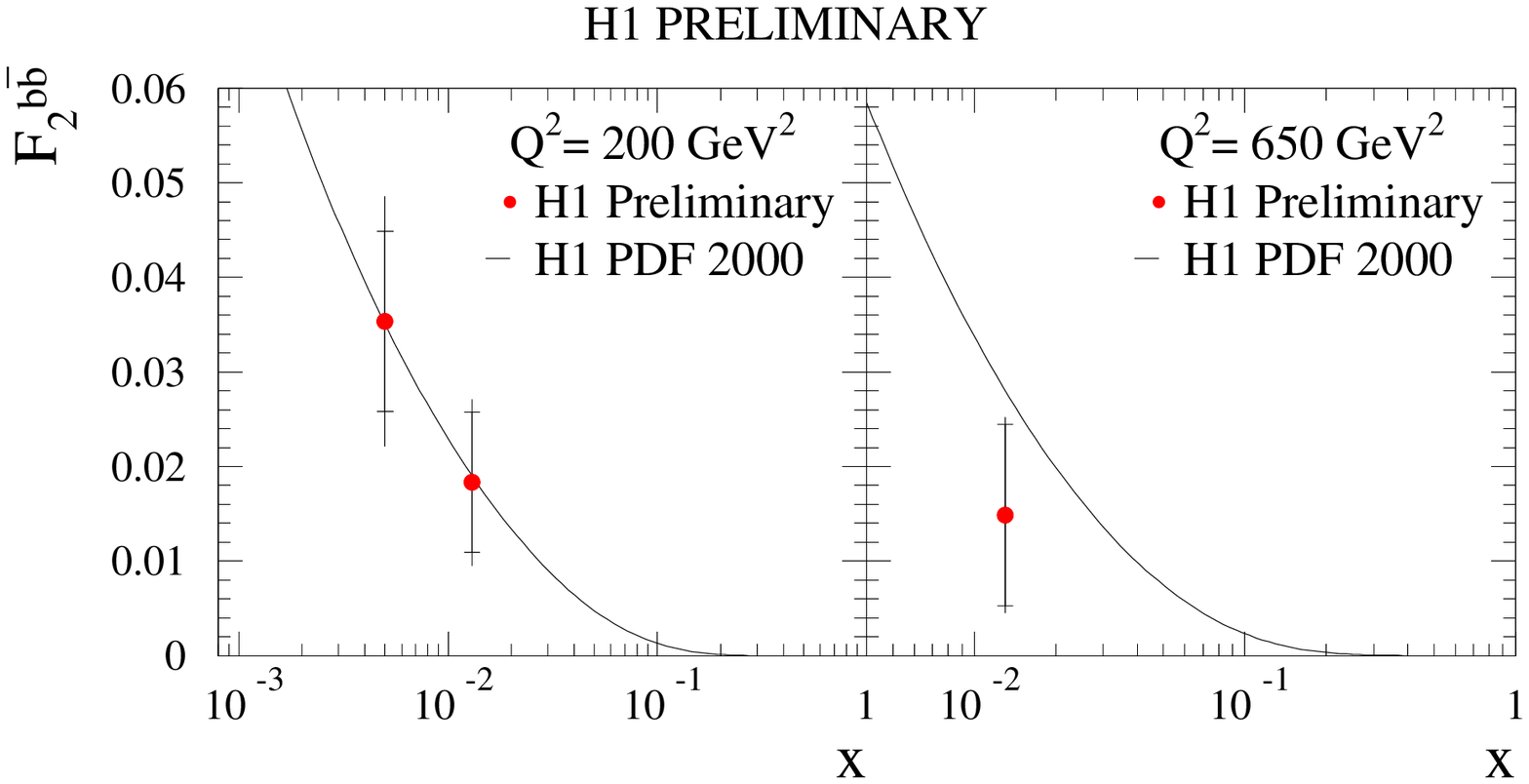}
\caption{ Measurements of the structure functions
  $F_2^{c\bar{c}}$ and $F_2^{b\bar{b}}$. The data are plotted as a
  function of $x$ for 2 values of $Q^2$.  Also included is
  the results of a NLO QCD fit to inclusive data. }
\label{fig:f2ccbb}

\end{figure}

There have been 2 methods employed at HERA to extract the charm and
beauty structure functions $F_2^{c\bar{c}}$ and $F_2^{b\bar{b}}$ of
the proton. The first method is to tag charm by reconstructing $D^*$
mesons.  This method results in a clean high statistics event sample
and has been used by H1 and ZEUS to measure $F_2^{c\bar{c}}$ over the
range $2<Q^2<500$~$\rm{GeV}^2$ \cite{H1ZEUSDstar}. The drawback of
this method is that only a small fraction of charm events are seen and
model dependent corrections must be employed to correct for unseen
phase space, particularly at low transverse momentum of the $D^*$, in
order to extract $F_2^{c\bar{c}}$. In the second method charm and
beauty events are distinguished from each other and from the light
quarks by measuring the displacement of tracks from the primary
vertex~\cite{h1f2ccf2bb}. The longer lived heavy hadrons produce
tracks with significant displacements from the primary vertex, while
the light hadrons usually decay so fast that the tracks are produced
at the primary vertex. This method is used to measure $F_2^{c\bar{c}}$
and to make the first measurement of $F_2^{b\bar{b}}$. The
measurements of the structure functions are shown in
figure~\ref{fig:f2ccbb} for the high $Q^2$ region. The 2 methods used
to extract $F_2^{c\bar{c}}$ are found to be in good agreement. The
measurements are also compared to the predictions of NLO QCD, which is
found to give a good description of both the charm and beauty
structure functions.

\end{document}